\documentclass[aps,prl,twocolumn,showpacs,superscriptaddress,amsmath,longbibliography,amssymb]{revtex4-1}

\usepackage[varg]{txfonts}   
\usepackage{hyperref}
\usepackage{url}
\usepackage{xcolor}  
\usepackage{graphicx}
\usepackage{color,soul}

\hypersetup{colorlinks=true,citecolor=blue,urlcolor=blue,linkcolor=blue}
\begin{document}
\title{Ranked percolation model for the caking time of amorphous molecular powders}

\author{Vasco C. Braz} \email{vcbraz@fc.ul.pt}
\affiliation{Centro de F\'{i}sica Te\'{o}rica e Computacional, Universidade de Lisboa, 1749-016
Lisboa, Portugal}
\affiliation{Departamento de F\'{\i}sica, Faculdade de Ci\^{e}ncias,
Universidade de Lisboa, 1749-016 Lisboa, Portugal}
  
\author{N. A. M. Ara\'ujo} \email{nmaraujo@fc.ul.pt}
\affiliation{Centro de F\'{i}sica Te\'{o}rica e Computacional, Universidade de Lisboa, 1749-016
Lisboa, Portugal}
\affiliation{Departamento de F\'{\i}sica, Faculdade de Ci\^{e}ncias,
Universidade de Lisboa, 1749-016 Lisboa, Portugal}  

\begin{abstract}
When amorphous molecular powders are exposed to high humidity levels or temperatures, the particle viscosity increases due to plasticization, promoting the formation of sinter bridges between pairs of particles in contact. Over time, these bridges facilitate particle agglomeration, eventually leading to the formation of a macroscopic cake that alters the mechanical properties and affects product quality. In this work, we model the caking process of amorphous powders subjected to a temperature shock as a bond percolation problem and investigate how particle bed heterogeneities influence the percolation threshold and, consequently, the expected caking times. Our findings indicate that a slight dispersion in particle size lowers the percolation threshold compared to a monodisperse bed or random percolation in polydisperse systems. Furthermore, we show that the expected caking time exhibits a non-monotonic behavior with the size dispersion, initially decreasing for low dispersion values and increasing for higher values. These results provide insights into the role of the particle size distribution on the caking dynamics of amorphous molecular powders.

\end{abstract}

\maketitle

\section{Introduction}
\label{intro}
The transition of free-flowing powders composed of molecular amorphous particles into “lumped” or “caked” states poses significant problems to their handling and perceived quality in a wide range of industries, from food products~\cite{Fitzpatrick2007,Masum2020,Carpin2017} to bulk chemicals~\cite{Cleaver2004}, fertilizers ~\cite{Albadarin2017}, and pharmaceuticals~\cite{Kristensen1987}. 
The caking process primarily involves two mechanisms~\cite{Palzer2005,Hartmann2011,Zafar2017,Chen2019}: moisture sorption (at the individual particle level) and bridge sintering (between particle pairs), each occurring on different timescales. During the aggregation process, mechanisms with different timescales compete, and, depending on the surrounding environmental conditions, their relative importance might change. 
Recently, it was demonstrated that the caking of amorphous powders can be mapped into a percolation problem where site occupation follows a well-determined order, depending on the physical causes that led to the caking process~\cite{Braz2022,Simoes2022}.
Up to now, only humidity shock scenarios were considered, where the typical sorption time is much larger than the sinter time. In this case, particles in contact form bridges instantly and, since the plasticization of individual particles is the result of water diffusion through the particle surface, the first particles to became become available for sintering are the smaller ones~\cite{Braz2022, braz2025}. From a fundamental perspective, caking due to humidity shock can be well described as a ranked site percolation process where particle size defines the order of occupation~\cite{Schrenk2013}. 
\begin{figure*}[t]
\centering
\includegraphics[width=1\textwidth]{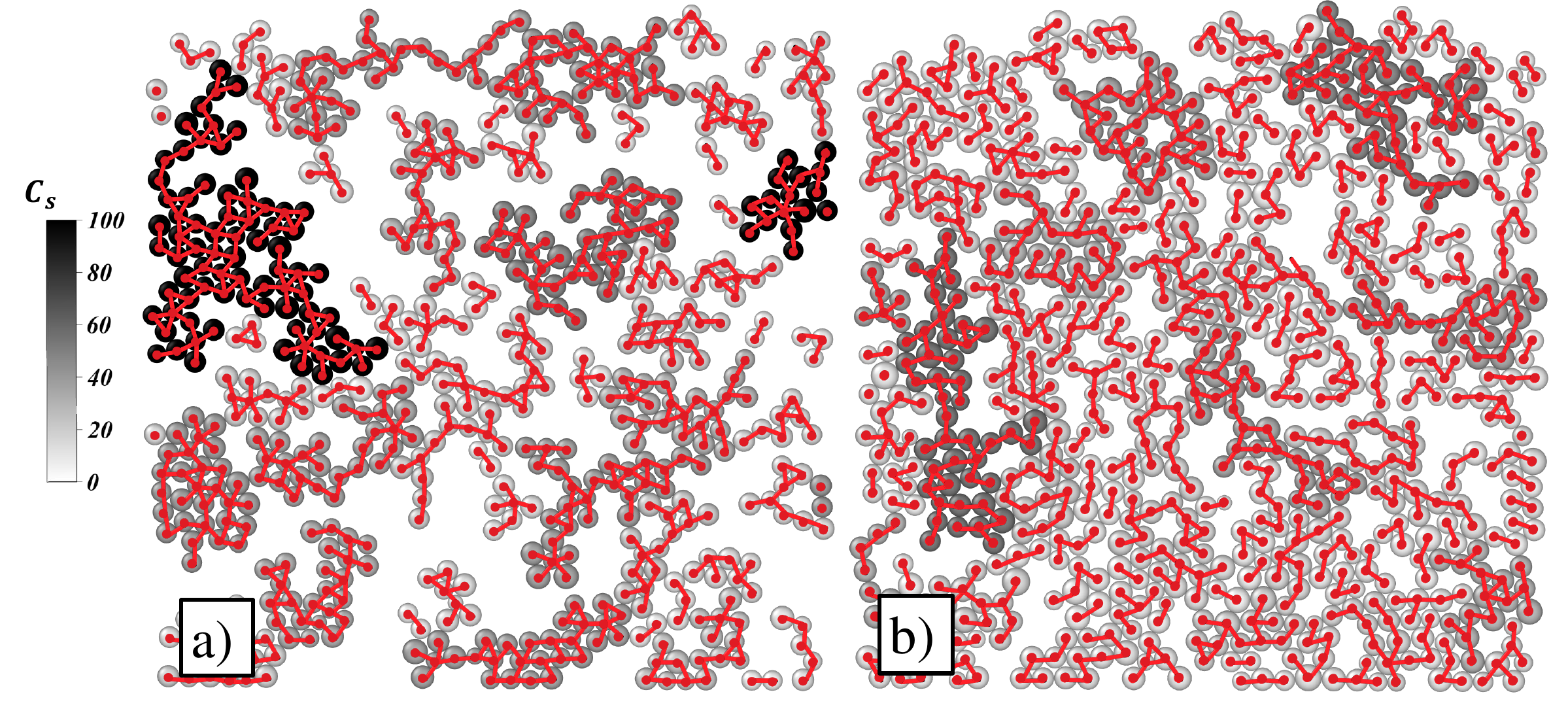}
\caption{Snapshot of a two-dimensional system with a bond occupation fraction $p=0.33$ and $\sigma=0.1$. Clusters are colored according to their size $C_s$ (number of particles in the cluster), ranging from light grey (small clusters) to black (largest cluster). The system has periodic boundary conditions along the horizontal direction. For clarity, bonds crossing boundaries are not shown, although clusters that span the boundary are still represented as connected (e.g., the black cluster in panel (a)).
(a) Size-Ranked Percolation.
(b) Random Percolation.}
\label{fig:snapshot}  
\end{figure*}
On the other hand, when powders are initially in moisture equilibrium with the environment and the temperature increases suddenly, the caking time is dominated by the sintering.  Here, the order of occupation is not determined solely by the individual particle size, but by the characteristics of the pairs of neighboring particle. The time required for sintering depends particle diameter ($d$)  and the ratio of surface tension $\gamma$ to viscosity ($\eta$), as described by the Frenkel equation~\cite{Hartmann2011}: 
\begin{equation}
   \left( \frac{x}{d} \right)^2 = \frac{\gamma t}{6 d \eta}, 
   \label{Eq:Frenkel}
\end{equation}
where, $x$ represents the diameter of the neck between the pair of particles.
In polydispersed particle beds, smaller particles in contact form bridges first, leading to particle aggregation in a specific sequence governed by the contact network structure. If all particles are of the same material, caking can be effectively modeled as a ranked bond percolation process where bond occupation follows a certain rank. The percolation threshold can then be related with the caking time~\cite{Braz2022}, which in the single species case, will be proportional to the typical size of the largest bridge ($d^*$) needed (if occupied in order of the average size of the pair). Assuming that a strong bridge is formed when $\left( \frac{x}{d} \right) = 0.1$~\cite{Wallack1988}, the caking time can be estimated from Eq.~(\ref{Eq:Frenkel}) as
\begin{equation}
t_c = 0.01 \frac{6 \eta d^*}{\gamma}.
\label{fgr:caking_time}
\end{equation}

We note that this same approach has been employed in numerical simulations to predict the caking times of various powders, yielding results that are in good agreement with experimental observations both in~\cite{Braz2022} and \cite{Simoes2022}.

In this work, we vary the particle size distributions and systematically determine the ranked bond percolation threshold across different particle configurations. We consider a normal distribution of particle sizes of mean $\mu=1$ and standard deviation $\sigma$ truncated between $r_l=0.1$ and $r_r=1.9$. For each $\sigma$ we generated 500 samples of 5000 particles. The particle beds are obtained by generating the spheres in a box without overlapping and and letting them sediment using a Discrete Element Method algorithm well described in~\cite{Braz2022}.

For comparison, three different strategies of occupation were used: S - Size Ranked Percolation, where bonds are occupied by the order of the average radius of pairs of particles in contact (Fig.~\ref{fig:snapshot}a), R - Random Percolation, bonds are occupied at random (Fig.~\ref{fig:snapshot}b), and D - Degree Ranked Percolation where occupation is done in the order of the average degree (number of contacts) of pairs of particles in contact.
\begin{figure}
\centering
\includegraphics[width=\columnwidth]{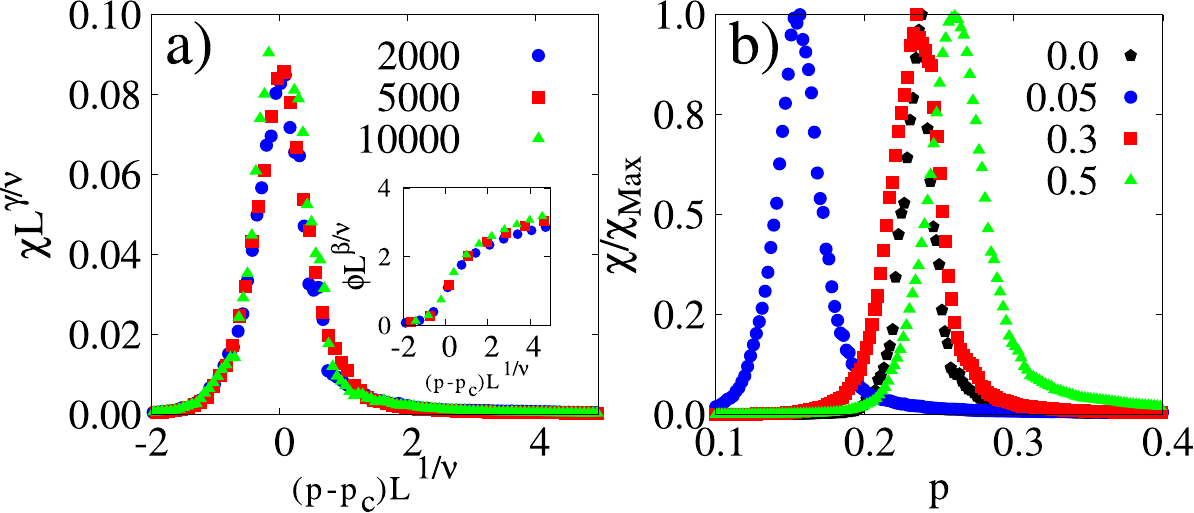}
\caption{(a) Rescaling of the fluctuations in the largest cluster size, using the typical exponents for three-dimensional percolation. The inset shows the same scaling but for the fraction of particles in the largest aggregate, $\phi$. We consider three different system sizes: $N=2000$ (blue circles), $N=5000$ (red squares), and $N=10000$ (green triangles). (b) Fluctuations in the largest cluster size as a function of the fraction of occupied sites for different size dispersions $\sigma = 0.0, 0.05, 0.3, 0.5$.}
\label{fig:percolation}       
\end{figure}
\section{Ranked bond percolation threshold}
\label{sec-1}
To estimate the percolation threshold, we measure the order parameter, defined as the average fraction of particles in the largest cluster, $\phi$, along with its fluctuations, $\chi = \langle \phi^2 \rangle - \langle \phi \rangle^2$~\cite{Araujo2010}, across 500 independent samples for each $\sigma$, as a function of the fraction of occupied bonds, $p$.

From Fig. ~\ref{fig:percolation}, it is clear that the Size Ranked Percolation follows the same scaling relation observed in random percolation in three dimensions. Specifically, for $\sigma = 0.2$, we obtained a data collapse for the finite-size scaling using the critical exponents of random percolation, namely, $\nu = 0.876$, $\beta = 0.4181$, and $\gamma = 1.8$~\cite{Stauffer1985,Sahimi1994,Araujo2014,Ziff2017} for both the fluctuations (main plot) and order parameter (inset). Here, $N = 2000, 5000, 10000$ is the number of particles, and $L = N^{1/3}$ is the system size. The observed collapse of the curves confirms that Ranked Percolation belongs to the same universality class as Random Percolation.

In Fig.~\ref{fig:percolation}b, we plot the normalized fluctuations, $\frac{\chi}{\chi_{max}}$, as a function of the fraction of occupied bonds $p$ for different values of the size dispersion $\sigma$.  In the monodisperse case ($\sigma = 0$), where all particles have the same size, occupation occurs randomly, and the bond percolation threshold coincides with that of standard Random Percolation, with $p_c \approx 0.241$, consistent with values reported in the literature\cite{Ziff2017}. For a small dispersion ($\sigma = 0.05$), we observe a notable decrease in the critical threshold. However, as $\sigma$ increases further, $p_c$ gradually rises, eventually exceeding the monodisperse value for $\sigma = 0.5$.

\begin{figure}[h]
\centering
\includegraphics[width=\columnwidth]{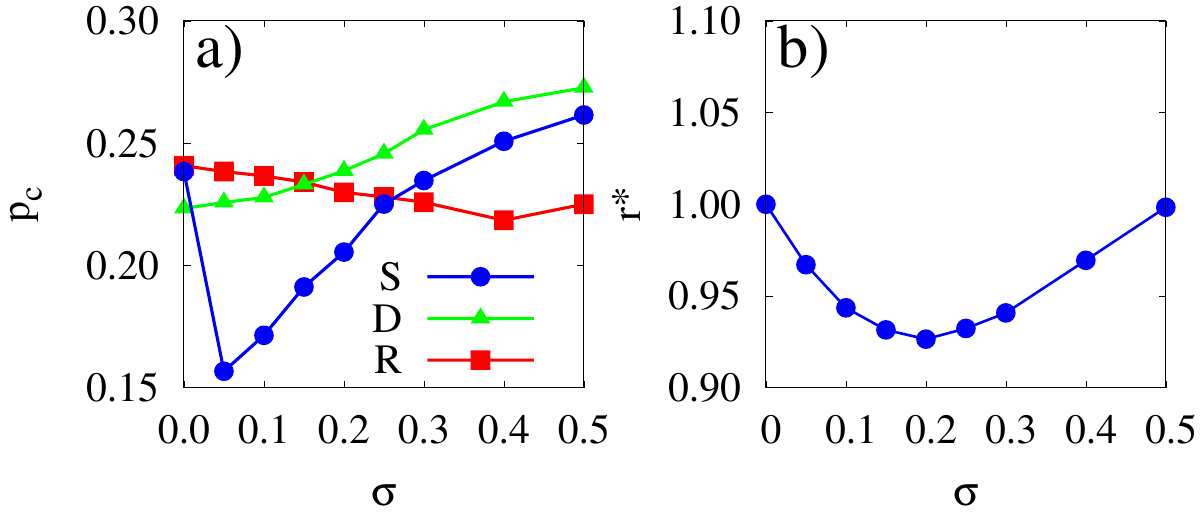}
\caption{(a) Percolation threshold as a function of size dispersion $\sigma$, for Size Ranked Percolation (S), Degree Ranked Percolation (D) and Random Percolation (R). (b) Maximum bridge radius at the percolation threshold for Size Ranked Percolation.}
\label{fig:threshold}       
\end{figure}

In Fig.~\ref{fig:threshold}a, we present a systematic study of the percolation threshold for Size Ranked Percolation (S – blue dots), Random Percolation (red squares), and Degree Ranked Percolation (D – green triangles) as a function of $\sigma$. As observed before for $\sigma = 0$, the percolation threshold matches that of standard Random Percolation. For small size dispersions (e.g., $\sigma = 0.05$), the occupation process is no longer random, leading to a significant drop in the percolation, which will remain lower than the values found for random percolation up to values of $\sigma=0.25$. As the size dispersion increases, the percolation threshold rises again. This intriguing reduction in percolation threshold under low size dispersion suggests that even minimal polydispersities introduces spatial and size correlations that enhance branching during size ranked bond occupation. Figure ~\ref{fig:snapshot}b illustrates this phenomenon in a 2D system: bonds are distributed uniformly in space, forming small clusters of isolated particles due to the limited size variation. By contrast, Fig.~\ref{fig:snapshot}b reveals how these correlations promote the early emergence of cycles (loops) but also branched, tree-like structures.  

As size dispersion increases, the coordination number depends strongly on the particle size (larger particles have greater surface area and thus more neighbors). Consequently, in Size Ranked Percolation, bonds connecting particles with fewer neighbors are occupied first: a behavior that qualitatively mirrors Degree Ranked Percolation, where occupation prioritizes low-coordination bonds.  Even in the absence of particle size dispersion, the disordered nature of the packing results in a distribution of coordination numbers, which is enough to affect the dynamics of degree-based percolation and lead to a higher percolation threshold relative to the random case. 

In Fig.~\ref{fig:threshold}b, we observe how the average diameter of the last particle pair to be connected before the percolation transition (value at $p_c$) evolves with $\sigma$. We emphasize that, according to the Frenkel equation, for a single species particle bed at fixed temperature and with fixed water content this diameter is proportional to the time it takes for that bridge to correspond to a rigid bond. Since this is the typical maximum average diameter at the percolation threshold, it serves as a proxy for the caking time. In Fig.~\ref{fig:threshold}b we show that the characteristic average radius for percolation $r^*$ first decreases with $\sigma$ in relation to the monodisperse case, a consequence of the broadening of the size dispersion and of the decrease in the $p_c$. This value keeps decreasing reaching a minimum at $\sigma=0.2$ and then it increases, matching the value of the monodispersed case at $\sigma=0.5$, the highest value simulated.  We highlight that, while the non-monotonic behavior of the percolation threshold arises due to the sudden emergence of correlations when a minimum size dispersion is introduced, above which the behavior becomes monotonic, the non-monotonicity observed for $r^*$ stems from the interplay between the overall increase of $p_c$ with $\sigma$  and the simultaneous appearance of a growing number of smaller particles as $\sigma$ increases. In short, more particles are needed at intermediate to high size dispersions, but they may, in fact, be smaller.
\section{Number of short cycles}
\begin{figure}
\centering
\includegraphics[width=\columnwidth]{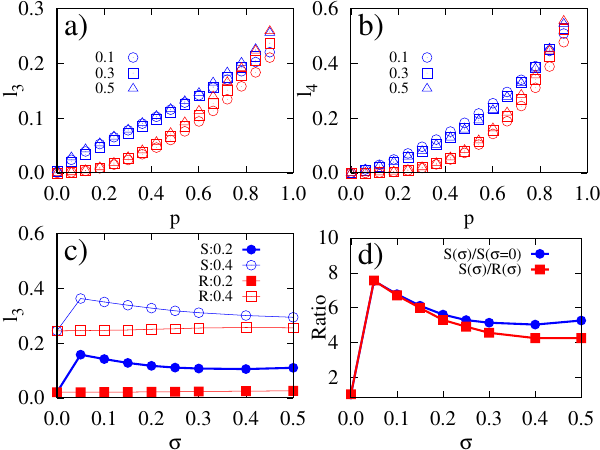}
\caption{(a) Fraction of loops of size 3 relative to the number of occupied bonds as a function of the fraction of occupied bonds.  
(b) Fraction of loops of size 4 relative to the number of occupied bonds as a function of the fraction of occupied bonds.  
(c) Fraction of loops of size 3 relative to the number of occupied bonds as a function of size dispersion $\sigma$.  
(d) In blue, the fraction of loops of size 3 in Size Ranked Percolation as a function of $\sigma$, normalized by its value at $\sigma = 0$. In red, the same quantity, but normalized by the corresponding values in Random Percolation.  
}
\label{fig:loops}       
\end{figure}

The mechanical stability of a granular pack is closely linked to the number of structural cycles within its network. Specifically, 3-cycles (triangular cycles) are known to correlate with stiffness. Figures ~\ref{fig:loops}a and ~\ref{fig:loops}b show the evolution of the average number of 3-cycles (\( l_3 \)) and 4-cycles (\( l_4 \)) as a function of the fraction of occupied bonds (\( p \)). The number of l-cycles is numerically calculated from the adjacency matrix and by counting the number of paths of length $l$, that start and end at the same node. For both cycle orders, the size-based bond occupation process (blue symbols) presents significantly more cycles at the same \( p \) compared to random bond occupation (red symbols). In Fig.~\ref{fig:loops}c we see that for fixed values of $p=0.2,0.4$, the number of cycles increases with $\sigma$ during random occupation but decreases for size ranked occupation. In Fig.~\ref{fig:loops}d we plot the ratio of $\frac{l^S_3(\sigma)}{l^S_3(\sigma=0)}$ (blue circles)  and  $\frac{l^S_3(\sigma)}{l^R_3(\sigma=0)}$ (red squares), the ratio between the number of cycles in size ranked percolation and random percolation as a function of $\sigma$. In both cases we observe, once again, the coincidence for the monodispersed pack, a sudden increase for low dispersion and a decrease for higher dispersions. 
\section{Conclusions}
In this study, we map the caking of amorphous molecular powders under sudden thermal shock to a ranked bond percolation problem, where bonds are occupied (or activated) in order of the average size of the corresponding pair of particles. Additionally, we quantify the impact of structural heterogeneity on the caking times of granular materials.  
   
We first determined the ranked bond percolation threshold as a function of the size dispersion of the pack. We found that for low dispersion values, the percolation threshold decreases in relation to the monodisperse packs (and random percolation on disperse packs), and then increases monotonically with the dispersion. The characteristic average radius at the percolation $p_c$ shows a non-monotonic behavior. Since the typical times for caking should be proportional to this characteristic length, we deduce that low size dispersions of the granular pack will lead to faster caking of the powder, while higher size dispersions will increase the expected caking time. 

Finally we measured the number of short cycles and show that Size Ranked Percolation will lead to a higher number of cycles than Random Percolation, showing a decrease with size dispersion. This finding suggests that not only might a spanning aggregate emerge first for low dispersed packs, but also that the spanning cluster might be more mechanically stable.

These results allow for a better understanding of the role of spatial and size correlations of the granular packing in the caking of amorphous powders. We emphasize that the work presented here still lacks experimental confrontation and that this should be explored in the future. 

 \bibliography{bib} 

\end{document}